\documentclass{article}

\PassOptionsToPackage{numbers, compress}{natbib}
\usepackage[final]{neurips_2024} 
\usepackage{adjustbox}
\usepackage[utf8]{inputenc} 
\usepackage[T1]{fontenc}    
\usepackage{hyperref}       
\usepackage{url}            
\usepackage{booktabs}       
\usepackage{amsfonts}       
\usepackage{nicefrac}       
\usepackage{microtype}      
\usepackage{xcolor}         

\usepackage{amsmath}
\usepackage{algorithm}
\usepackage{algorithmic}
\usepackage{graphicx}
\usepackage{tcolorbox}
\usepackage{multirow}
\usepackage{svg}
\usepackage{enumitem}

\title{CPP-UT-Bench: Can LLMs Write Complex Unit Tests in C++?}

\author{%
  Vaishnavi Bhargava$^{1\thanks{Work done during internship at Nutanix}}$, Rajat Ghosh$^2$, Debojyoti Dutta$^2$ \\
  $^1$University of Wisconsin-Madison, $^2$Nutanix \\
  \texttt{vbhargava3@wisc.edu, \{rajat.ghosh, debojyoti.dutta\}@nutanix.com} \\
}

\begin{document}

\maketitle

\begin{abstract}

We introduce CPP-UT-Bench, a benchmark dataset to measure C++ unit test generation capability of a large language model (LLM). CPP-UT-Bench aims to reflect a broad and diverse set of C++ codebases found in the real world. The dataset includes 2,653 \{code, unit test\} pairs drawn from 14 different opensource C++ codebases spanned across nine diverse domains including machine learning, software testing, parsing, standard input-output, data engineering, logging, complete expression evaluation, key value storage, and server protocols. We demonstrated the effectiveness of CPP-UT-Bench as a benchmark dataset through extensive experiments in in-context learning, parameter-efficient fine-tuning (PEFT), and full-parameter fine-tuning. We also discussed the challenges of the dataset compilation and insights we learned from in-context learning and fine-tuning experiments. Besides the CPP-UT-Bench dataset and data compilation code, we are also offering the fine-tuned model weights for further research. For nine out of ten experiments, our fine-tuned LLMs outperformed the corresponding base models by an average of more than 70\%.

\end{abstract}

\section{Introduction}

Large Language Models (LLMs) \citep{dubey2024llama} have demonstrated impressive performance on a number of recently proposed coding benchmarks such as HumanEval \citep{chen2021evaluating}, MBPP, \citep{austin2021program}, and MultiPL-E \citep{cassano2023multipl}. Nonetheless, existing benchmarks, in general, have reached saturation \citep{kiela2021dynabench, ott2022mapping} and lack representation from real-world software engineering tasks \citep{srivastava2022beyond}. Evaluating coding performance on short and self-contained algorithmic tasks, existing coding benchmarks such as MBPP are far from the real-world software engineering tasks such as unit test writing. Moreover, the existing coding benchmarks mostly cover high-level languages such as Python. Lower-level languages (e.g., C, C++) have higher Kolmogorov complexity \citep{li2008introduction} and cyclomatic complexity \citep{lopes2022and} due to its verbosity, advanced features (e.g., templates, macros), and manual memory management. Therefore, a C++ codebase is harder to maintain and stands to benefit considerably from unit test generation automation. However, there is hardly any benchmark dataset for C++ unittest generation representative of the real world software engineering.

Inspired by this challenge of the lack of C++ unit test generation benchmark dataset, we introduce CPP-UT-Bench from diverse domains. We evaluate multiple state-of-the-art LLMs on CPP-UT-Bench and study their performances for few-shot in-context learning, parameter-efficient fine-tuning (PEFT), and full-parameter fine-tuning.

\section{CPP-UT-Bench}

CPP-UT-Bench is a benchmark featuring 2,653 \{code, unittest\} pairs from from 14 popular opensource C++ repositories with permissible licenses. CPP-UT-Bench is organized in the following schema:\footnote{Huggingface link to CPP-UT-Bench dataset: \url{https://huggingface.co/datasets/Nutanix/CPP-UNITTEST-BENCH}}
\footnote{Huggingface link to fine-tuning dataset: \url{https://huggingface.co/datasets/Nutanix/cpp_train_dataset_chat_format_less_than_8k}}

\begin{itemize}
    \item \textbf{ID}: A unique identifier for each entry in the dataset. [Example: "0"]
    \item \textbf{Language}: The programming language of the file. [Example: "cpp"]
    \item \textbf{Repository Name}: The name of the GitHub repository, formatted as organisation/repository. [Example: "google/googletest"]
    \item \textbf{File Name}: The base name of the file (without extension) where the code or test is located. [Example: "sample1"]
    \item \textbf{File Path in Repository}: The relative path to the file within the GitHub repository. [Example: "googletest/samples/sample1.cc"]
    \item \textbf{File Path for Unit Test}: The relative path to the unit test file, if applicable. [Example: "googletest/samples/sample1\_unittest.cc"]
    \item \textbf{Code}: The code content of the file, excluding any documentation or comments.
    \item \textbf{Unit Test (Ground Truth)}: The content of the unit test file that tests the code.
\end{itemize}

\begin{figure}[h!]
    \centering    \fbox{\includegraphics[width=0.8\textwidth]{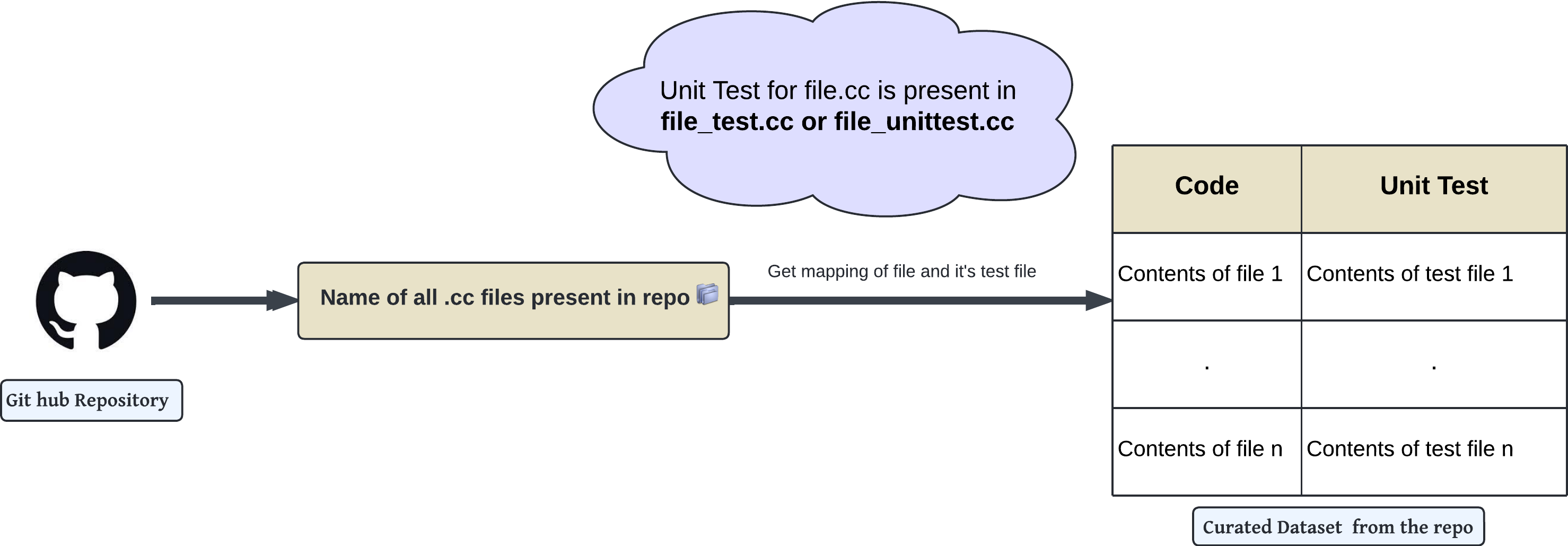}}
    \caption{Data extraction pipeline for CPP-UT-Bench. It uses GitHub repos as upstream sources and then processes the {code, unittest} pairs extracted to create the benchmark dataset.   }
    \label{fig:dataextraction}
\end{figure}

We collected this data from GitHub. Although GitHub is a rich data source for software engineering, not all codebases have sufficient unit test coverage. Also, the relationship between code and unit test is often noisy, ad-hoc, and poorly documented. Our data curation pipeline is designed to be generic and adaptable, making it applicable to diverse C++ codebases. To compile a high-quality C++ unit test generation benchmark at scale, we use the following two-step pipeline, as shown in Figure \ref{fig:dataextraction}. \footnote{Python Script to create CPP-UT-Bench dataset: \url{https://huggingface.co/datasets/Nutanix/CPP-UNITTEST-BENCH/blob/main/data_scrape.py}}

\begin{itemize}[left=0pt]
    \item \textbf{File Extraction and Grouping:} The initial phase involves extracting relevant files from the codebase. We concentrate on C++ source files (with extensions \texttt{.cc} and \texttt{.h}) and unit test files (with extensions \texttt{\_test.cc} and \texttt{\_unittest.cc}). A recursive directory search ensures comprehensive identification of these files. Once extracted, files are grouped by their base names, derived by stripping file extensions. For example, files named \texttt{Foo.cc} and \texttt{Foo.h} are grouped under the base name \texttt{Foo}, linking implementation files with their corresponding declarations. Similarly, test files are associated with their source files based on these shared base names.     
    \item \textbf{Mapping Source Files to Test Files and Documentation:} Following the extraction and grouping of C++ source files and unit test files, we map each source file to its respective test files. When both \texttt{\_test.cc} and \texttt{\_unittest.cc} files are present, we prioritize \texttt{\_test.cc}. This structured mapping is crucial for analyzing code coverage and evaluating the effectiveness of unit tests. The final stage of the process involves documenting the extracted data. For each base name, we compile detailed records of the repository name, source code content, and test code content into an Excel spreadsheet. This organized documentation enables comprehensive analysis, providing valuable insights into code coverage and the adequacy of unit tests.
    
\end{itemize}

\subsection{Task Formulation}

We evaluate CPP-UT-Bench for different tasks, as follows: 

\textbf{Few-Shot In-Context Learning}: Few-shot in-context learning (FS-ICL) in this work refers to the setting where the model is given a few demonstrations of the task at inference time as conditioning \citep{brown2020language}, but no weight updates are allowed. As shown in Equation \ref{fewshot_icl}, FS-ICL takes an query, $x_{test}$ at inference time and uses a fixed-parameter model, $f_{\theta}$, along with $k$ demonstrations, $(x_i, y_i)\}_{i=1}^{k}$, to produce a response, $y_{test}$. The response quality depends on the concerned LLM $f_{\theta}$ and the demonstration set. 

\begin{equation}
\label{fewshot_icl}
y_{\text{test}} = f_{\theta} \left( \{(x_i, y_i)\}_{i=1}^{k}, x_{\text{test}} \right)
\end{equation}

\textbf{Parameter-Efficient Fine-Tuning}: Parameter-efficient fine-tuning (PEFT) involves updating some subsets of weights of a pre-trained model, $f_{\theta}$ by training on a supervised dataset specific to a desired task. In general, at least a few thousands of labeled examples are used. While fine-tuning improves task-specific performances, it needs a large demonstration dataset. For PEFT, low-rank adaptation (LoRA) \citep{hu2021lora} is one of the most prevalent techniques, as shown in  Equation \ref{eq:lora}. 

\begin{equation}
\label{eq:lora}
\max_{\Theta} \sum_{(x,y) \in \mathbb{Z}} \sum_{t=1}^{|y|} \log \left( p_{\Phi_0 + \Delta \Phi(\Theta)} \left( y_t \mid x, y_{<t} \right) \right)
\end{equation}

\textbf{Full-Parameter Fine-Tuning}: Full-parameter fine-tuning \citep{lv2024parameterfinetuninglargelanguage} involves updating the all weights of a pre-trained model, $f_{\theta}$ by training on a supervised dataset specific to a desired task. Because it is updating all the weights, it comes with much higher computational cost than PEFT/LoRA. 

\begin{equation}
\label{eq:full-parameter-fine-tuning}
\max_{\Phi} \sum_{(x,y) \in \mathbb{Z}} \sum_{t=1}^{|y|} \log \left( P_{\Phi} \left( y_t \mid x, y_{<t} \right) \right)
\end{equation}

Both PEFT and full-parameter fine-tuning are inter-related. A pre-trained LLM, $P_\Phi(y \mid x)$ is parameterized by $\Phi$. A downstream task is represented by a training dataset of context-target pairs: $Z = \{(x_i, y_i)\}_{i=1,\dots,N}
$ where both $x_i$ and $y_i$ are sequences of tokens. During full fine-tuning, the model is initialized to the base weights $\Phi_0$ and updated to $\Phi_0 + \Delta \Phi$ by repeatedly following the gradient to maximize the conditional language modeling objective as shown in Equation \ref{eq:full-parameter-fine-tuning}. Fine-tuning the entire pre-trained weight space could be prohibitively expensive. That is where PEFT brings value. PEFT adopts a more parameter-efficient approach, where the task-specific parameter increment $\Delta \Phi = \Delta \Phi(\Theta)$ is further encoded by a much smaller-sized set of parameters $\Theta$ with $|\Theta| \ll |\Phi_0|$. The task of finding $\Delta \Phi$ thus becomes optimizing over $\Theta$, as shown in Equation \ref{eq:lora}.

\subsection{Features of CPP-UT-Bench}

\begin{figure}[h!]
    \centering    \fbox{\includegraphics[width=0.8\textwidth]{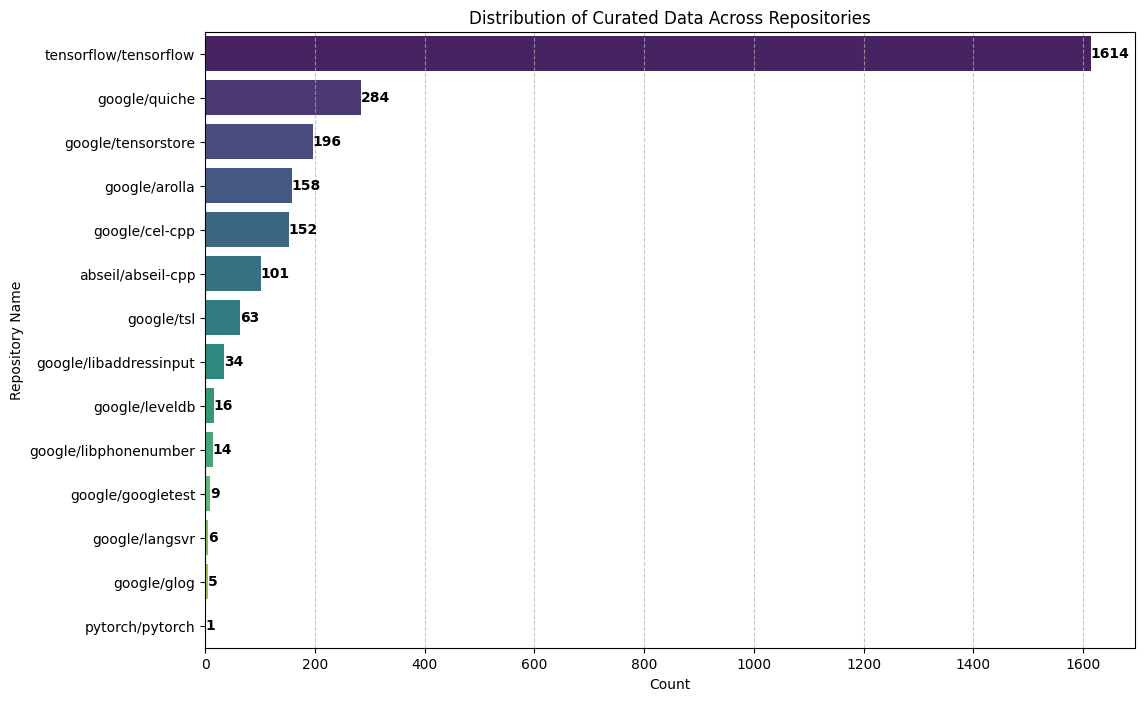}}
    \caption{Data distribution of CPP-UT-Bench from 14 different GitHub Repositories. The dominant contribution (greater than $60 \%$)  comes from Tensorflow and the least from PyTorch.  }
    \label{fig:cpp_ut_bench_dist}
\end{figure}

Traditional code benchmarks such as MBPP typically involve only short and standalone input and output sequences. In contrast, CPP-UT-Bench represents real-world software engineering in C++. CPP-UT-Bench consists of widely popular open-source code bases such as TensorFlow.  Figure \ref{fig:cpp_ut_bench_dist} shows the distribution of CPP-UT-Bench in terms of source repositories. It shows the dataset has imbalanced representation from different projects with the dominant being Tensorflow. Overall, it has 2,653 pairs from 14 open-source projects with permissible licenses covering nine different domains.

The domain diversity of CPP-UT-Bench is shown in Table \ref{tab:cpp_ut_bench_domain}. It covers a wide gamut of real world software applications including machine learning, data engineering, software testing, telecommunications, key-value storage, server protocol, geolocation, concurrency, and application logging. 
        
\renewcommand{\arraystretch}{1.5}
\label{tab:cpp_ut_bench_domain}
\begin{table}[ht]
\centering
\begin{tabular}{ll}
\hline
\textbf{Domain} & \textbf{Repository} \\ 
\hline
Machine learning  & Pytorch \citep{pytorch}, TensorFlow \citep{tensorflow}  \\ 
\hline
Storage and data engineering  & Tensorstore \citep{tensorstore} \\ 
\hline
Software testing  & Google Test \citep{google-test}, Abseil \citep{Abseil}, \\ \hline
Telecommunications & Libphonenumber \citep{libphonenumber} \\ \hline
Key-value storage  & LevelDB \citep{leveldb} \\ \hline
Server protocol & Langsvr \citep{langsvr}, Cel-cpp \citep{cel-cpp} \\ \hline
Geolocation  & Libaddressinput \citep{libaddressinput} \\ \hline
Concurrency and multi-threading & tsl \citep{tsl} \\ \hline
Application logging & glob \citep{glog} \\ \hline
\end{tabular}
\vspace{0.1cm} 
\caption{Domain diversity of CPP-UT-Bench.}
\end{table}

\begin{figure}[h!]
    \centering    \fbox{\includegraphics[width=0.8\textwidth]{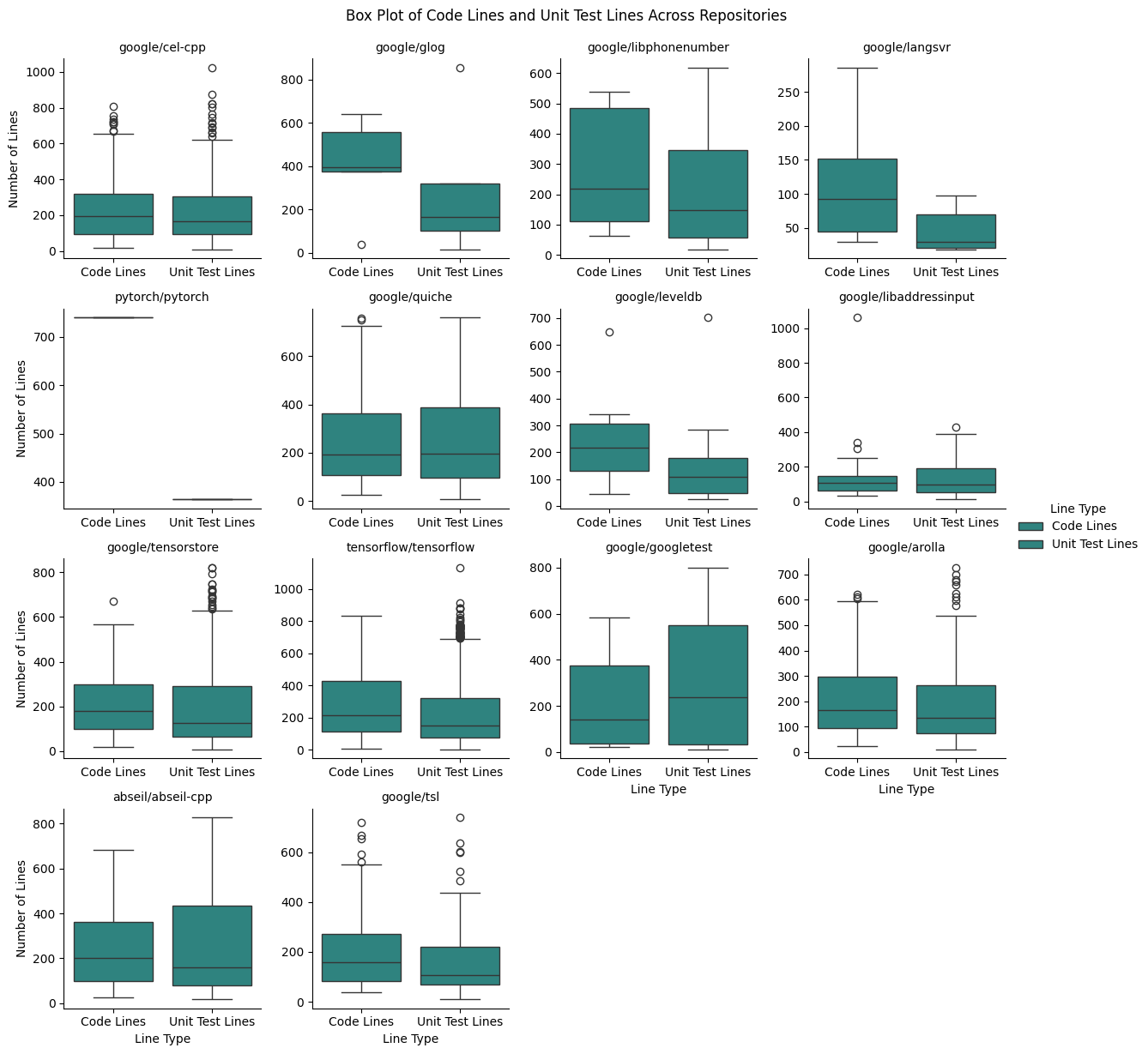}}
    \caption{The diversity in \{code, unit test\} pairs in terms of line lengths across 14 different opensource repositories in CPP-UT-Bench.  }
    \label{fig:relative_loc_diversity}
\end{figure}

The distribution of lengths for \{code, unit test\} pairs in CPP-UT-Bench grouped by different repositories is shown in Figure \ref{fig:relative_loc_diversity}. This shows all repositories have average line lengths greater than 100 with considerable variance and outliers, which is representative of the real-world code bases.

\subsection{LLM-as-a-Judge}\label{evaluation}

\begin{figure}[ht!]
    \centering    
\fbox{\includegraphics[width=\textwidth]{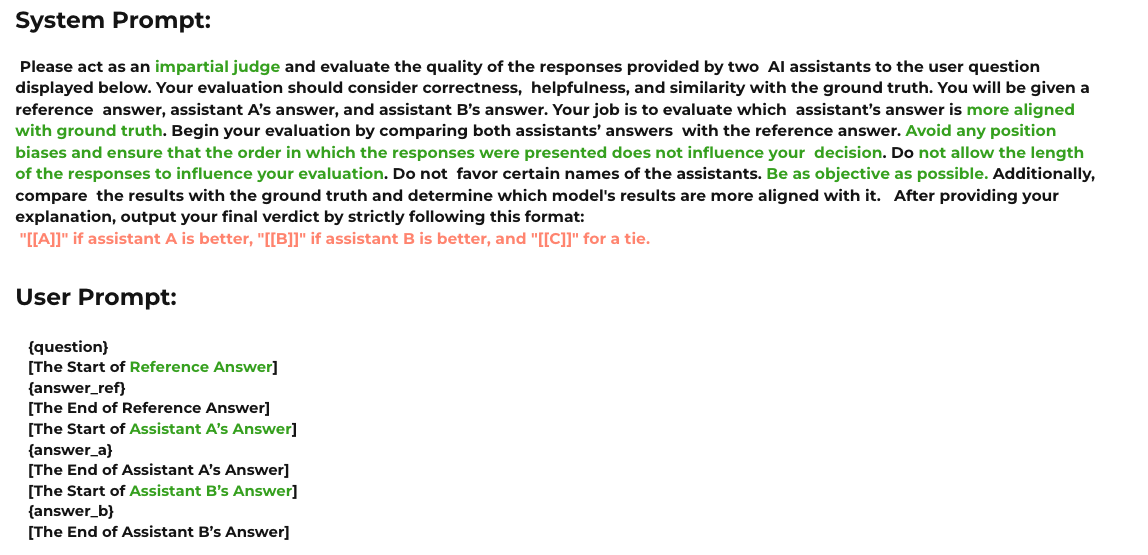}}
    \caption{Prompt for pairwise evaluation of two LLM generated responses (Assistant A and Assistant B) w.r.t. the ground truth.}
    \label{fig:evalprompt}
\end{figure}

To evaluate LLM performances in  few-shot in-context learning and fine-tuning, we have adopted LLM-as-a-Judge paradigm \citep{zheng2023judging} with GPT-4o-mini as the oracle model. This choice is to avoid the shortcomings of conventional NLP metrics such as BLEU and ROUGE which fail to effectively capture the semantic similarity required for evaluating generated code \citep{chen2021evaluating}. Equation \ref{eq:llm-as-a-judge} formally describes the standardized evaluation model, $\mathcal{E}$, we follow. It evaluates a triplet, $(r_A, r_B, g)$. $r_A$ is the response from LLM-A. $r_B$ is the response from LLM-B. $g$ is the ground truth. The oracle LLM judges between $\{r_A, r_B\}$ which response is more closely aligned to $g$.  The alignment judgement function, $J$ and the relative comparison between two alignments is executed by the Oracle LLM itself in a zero-shot manner with the prompt template shown in Figure \ref{fig:evalprompt}. The evaluation prompt was carefully designed to capture subtle differences between the outputs of the models and their alignment with the ground truth. To mitigate potential biases, such as GPT’s preference for longer responses or positional bias, we further tuned the prompt.

In this pairwise approach, the win rate for an evaluation set is defined as the percentage of instances where first model’s output is judged to be more closely aligned with the ground truth compared to the competing second model’s output. This method is supported by numerous studies demonstrating that GPT-based evaluation closely mimics human judgment while being less expensive and time-consuming \citep{zheng2023judging}. 

 
.

\begin{equation}
\mathcal{E}(r_A, r_B, g) =
\begin{cases}
r_A & \text{if } J(r_A, g) > J(r_B, g) \\
r_B & \text{if } J(r_A, g) < J(r_B, g) \\
\text{Tie} & \text{if } J(r_A, g) = J(r_B, g)
\end{cases}
\label{eq:llm-as-a-judge}
\end{equation}


\subsection{Framework for C++ Unit Tests Generation}\label{pipeline}

We use the following three-step workflow to generate the unit test given a source file. This is an important design consideration for our work given that many of the real world C++ code-base often exceed the context length for an LLM. 


\begin{enumerate}[left=0pt]
\item \textbf{Code Chunker}: In scenarios where a C++ class file exceeds 200 lines, it becomes suboptimal to prompt the LLM to generate unit tests for the entire file in one go. To address this, we implemented a method that processes the code file by generating multiple smaller chunks.  We leveraged the code chunker introduced by SweepAI which employ Concrete Syntax Tree (CST) based strategies \citep{CST}. Equation \ref{eq:cst} describes CST based chunking formally with $T(r)$ is the CST for the code $r$ and $C_{i}$ is the $i^{th}$ code chuck. It's designed to handle extremely large files by breaking them down into manageable sections that preserve the code's structure and context. This ensures each code chunk remains coherent and contextually relevant, thereby improving the accuracy and reliability of the generated unit tests.

\begin{equation}
T(r) = \{C_1, C_2, \dots, C_n\}
\label{eq:cst}
\end{equation}

\item \textbf{Unit Test Generation for a chunk}: 
For each chunk $i$, we prompt the LLM to generate the corresponding unit test, ${ICL}(C_i)$. Through extensive prompt engineering, we have refined the prompts to achieve optimal results. The prompt template is shown in Appendix, Figure \ref{fig:prompt template}.

\begin{equation}
UT(C_i) = \{\text{ICL}(C_i)\}
\end{equation}

\item \textbf{Compilation of unit test chunks}: Finally, we take the generated unit tests for the chunks and simply append them to give the final unit test file. This can further be enhanced by having an LLM prompt for combining the unit tests. 

\begin{equation}
UT(T(r)) = \sum_{i=1}^{n}UT(C_i)
\end{equation}
\end{enumerate}

\section{Experiment Design}
\label{sec: exp des}

The key value of a benchmark dataset such as CPP-UT-Bench comes from its value as a test data for few-shot in-context and a demonstration dataset for PEFT and full-parameter fine-tuning.

\textbf{Research Question-1 (RQ-1)}: Can CPP-UT-Bench replicate known results from well-known benchmarks in few-shot in-context learning? To answer this question, we compare the two-shot in-context learning performances for the following pairs: \{Phi-3-medium \citep{phi-3-medium} vs Phi-3-Small \citep{phi-3-small}\}, \{Mistral-7B-Instruct-v0.2 \citep{mistral-v0.2} vs  Mistral-7B-Instruct-v0.1 \citep{mistral-v0.1} \}, and \{Llama-3-70B-instruct-awq \citep{llama-70b-awq} vs Llama-3-8B-instruct-awq \citep{llama-8b-awq} \}. 

To evaluate the two-shot performance across these models, we employed the pipeline described in Section \ref{pipeline}, to generate unit tests for various code files. For inference, we configured the sampling parameters uniformly across both the original and fine-tuned models, setting a temperature of 0.1, a maximum token limit of 4,096, a frequency penalty of 0.3, and a top-p value of 0.7. We conducted preliminary experiments with various parameter values to determine these optimal settings.

\begin{figure}[h!]
    \centering    \fbox{\includegraphics[width=0.8\textwidth]{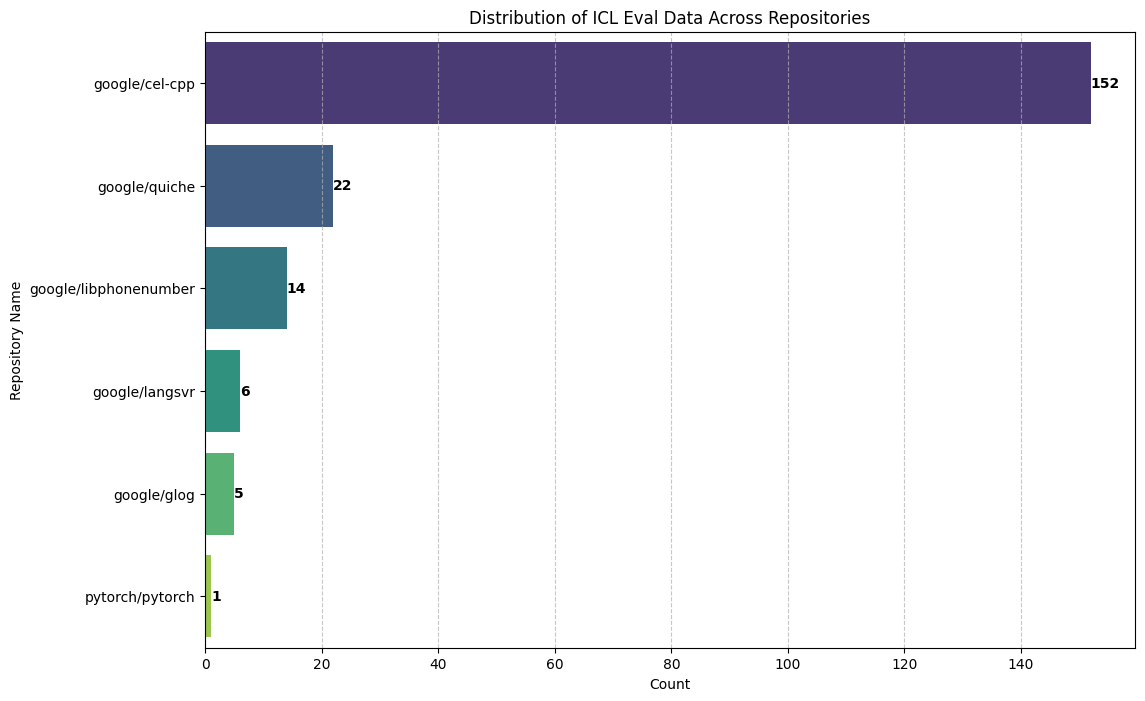}}
    \caption{Distribution of evaluation dataset for the few-shot in-context learning.} 
    \label{fig:icl_eval_dist}
\end{figure}

We generated unit tests for each model using 200 samples from the evaluation dataset, and then applied the methodology from Section \ref{evaluation} to assess model performance.  The evaluation was conducted using \href{https://openai.com/index/gpt-4o-mini-advancing-cost-efficient-intelligence/}{GPT-4o-mini} as the judge, and the comparison was quantified through win rate. Figure \ref{fig:icl_eval_dist} shows the distribution of evaluation data used for the few-shot in-context learning experiments. The repository choice has been somewhat random. In future, we will perform more analysis for other data distributions.  

\textbf{Research Question-2 (RQ-2)}: Does a full-parameter fine-tuned LLM with CPP-UT-Bench dataset performs better than its PEFT counterpart relative to a base LLM? To answer this question, we have compared both PEFT and full-parameter fine-tuned versions w.r.t. the corresponding base versions for five LLMs, including Mistral-7B-Instruct-v0.2 \citep{mistral-v0.2}, TinyLlama-1.1B-Chat-v1.0 \citep{tinyllama-1.1b}, CodeLlama-7B-Instruct \citep{codellama-7b}, Llama-3-8B-Instruct \citep{Llama-3-8B}, and Llama-3.1-8B-Instruct \citep{Llama-3.1-8B}.

\begin{itemize}[left=0pt]
    \item \textbf{PEFT Finetuning: } For our fine-tuning experiments, we used the Low-Rank Adaptation (LoRA) technique. Through a grid search, we optimized the LoRA parameters and found that a rank of 8 and an alpha of 16 yielded the best results. The fine-tuning was performed over two epochs on our curated dataset, with a learning rate of $5\times10^{-5}$.  We observed that using a smaller learning rate led to more stable training. LoRA was applied to the dense layers, including the gate\_proj, down\_proj, and up\_proj layers of the MLP block, as well as the q\_proj, v\_proj, k\_proj, and o\_proj layers in the Attention block. These layers provided the most effective results during training. The detailed hyper-parameter choices for the fine-tuning experiments are shown in Appedix (Table \ref{tab:fine_tuning_hyperparameters}).

    \item  \textbf{Full-Parameter Finetuning: }For the full fine-tuning or domain adaptation approach, we fine-tuned all the parameters of the model. We trained for two epochs on our dataset, using a learning rate of $5 \times10^{-5}$.
\end{itemize}

\begin{figure}[h!]
    \centering    \fbox{\includegraphics[width=0.8\textwidth]{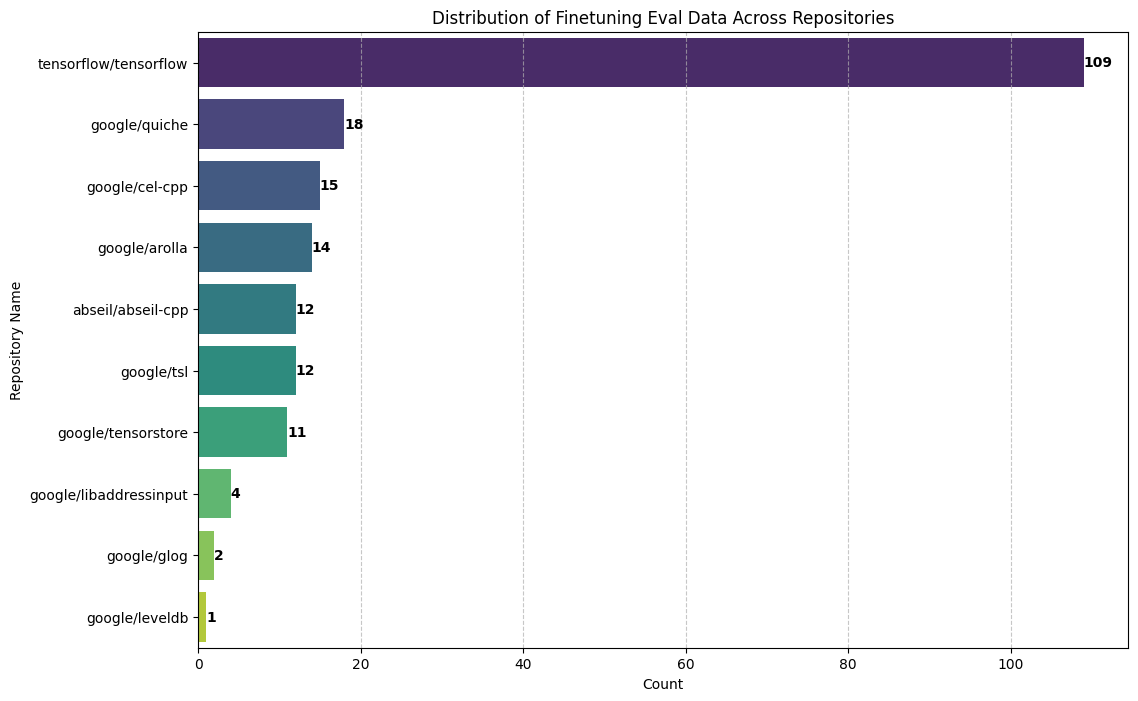}}
    \caption{Distribution of evaluation dataset for the fine-tuning experiments.} 
    \label{fig:ft_eval_dist}
\end{figure}

To evaluate the performance of the fine-tuned models against their original counterparts, we used the process mentioned in RQ-1. We employed the same pipeline (Section \ref{pipeline}) and sampling parameters for inference, and the results were evaluated using the methodology in Section \ref{evaluation}, with \href{https://openai.com/index/gpt-4o-mini-advancing-cost-efficient-intelligence/}{GPT-4o-mini} as the judge, quantifying performance via win rate. Figure \ref{fig:ft_eval_dist} shows the distribution of evaluation dataset for the fine-tuning experiments. The repository choice has been somewhat random. In future, we will perform a thorough ablation study. 

\section{Results}
This section is divided into two sub-sections each for two research questions. 

\subsection{Results for Few-Shot In-Context Learning (RQ-1)}

\begin{figure}[h!]
    \centering    \fbox{\includegraphics[width=0.8\textwidth]{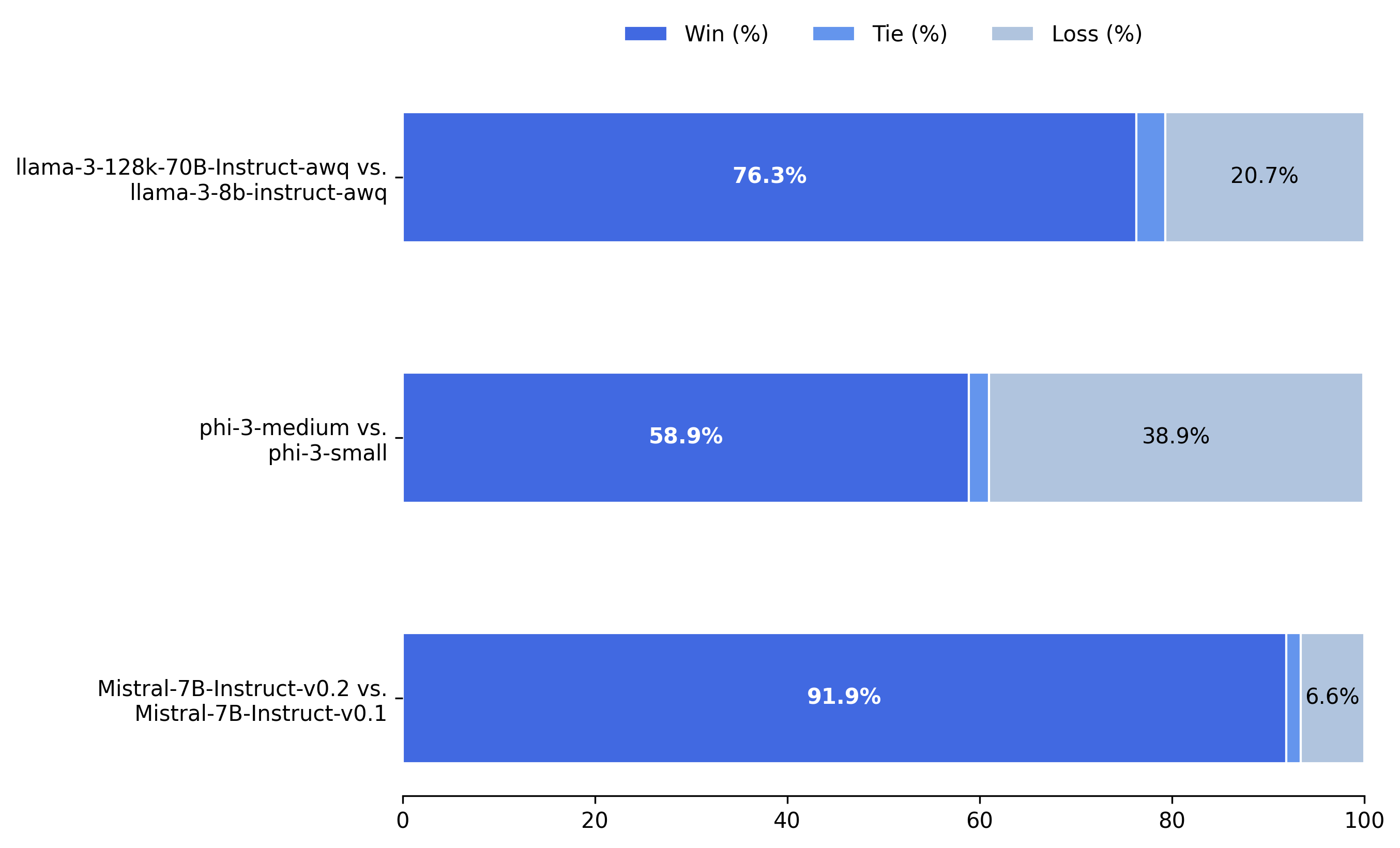}}
    \caption{Few-Shot in-context learning performance assessment for three LLM pairs: \{Llama-3-70B-instruct-awq \citep{llama-70b-awq} vs Llama-3-8B-instruct-awq \citep{llama-8b-awq} \}, \{Phi-3-medium \citep{phi-3-medium} vs Phi-3-Small \citep{phi-3-small}\}, and \{Mistral-7B-Instruct-v0.2 \citep{mistral-v0.2} vs  Mistral-7B-Instruct-v0.1 \citep{mistral-v0.1} \}. The results corroborate with other general coding benchmarks \citep{abdin2024phi, mistral-v0.2, dubey2024llama, jiang2023mistral}. } 
    \label{fig:icl_benchmarking}
\end{figure}

In few-shot in-context learning, we accessed the performance of three LLM families: Llama-3, Phi-3, and Mistral-7B-v0.2, as shown in Figure \ref{fig:icl_benchmarking}. 

\textbf{Llama-3 Family}: From Figure \ref{fig:icl_benchmarking} (top), we see Llama-3-128K-70B is winning over Llama-3-8B 76.3\% times. This can be attributed to higher context length and longer context length of Llama-3-128K-70B. This also corroborates the existing benchmarks \citep{dubey2024llama}.

\textbf{Phi-3 Family}: From Figure \ref{fig:icl_benchmarking} (mid), we see Phi-3-medium is winning over Phi-3-small 58.9\% times. Phi-3-small of 7B parameters and Phi-3-medium of 14B parameters are both trained for 4.8T tokens. They perform respectively 75\%, 78\% on MMLU, and 8.7, 8.9 on MT-bench \citep{abdin2024phi}. Following similar trends, our result also show slightly superior performance for Phi-3-medium. 

\textbf{Mistral-7B Family}: From Figure \ref{fig:icl_benchmarking} (bottom), we see Mistral-7B-Instruct-v0.2 is winning over Mistral-7B-Instruct-v0.1 a whopping 91.9\% times. Although both models have same parameter counts, Mistral-7B-Instruct-v0.2 several important characteristics that have possibly contributed to its superiority in C++ unit test generation. First, one of the most significant upgrades in v0.2 \citep{mistral-v0.1, mistral-v0.2} is the increase in the context window from 8k to 32k tokens. This allows the model to handle and generate longer sequences more efficiently, improving its ability to maintain context in larger inputs, especially for complex C++ unit test generation tasks. Second, the positional encoding mechanism was fine-tuned in v0.2, with the Rope-theta parameter adjusted to $10^{6}$. This optimization allows better handling of longer token sequences in C++ unit test generation. Finally, v0.2 drops the use of sliding window attention, a mechanism used in v0.1, which limits the model’s ability to capture long-range dependencies. By eliminating this feature, v0.2 improves its understanding of full input sequences, possibly contributing to enhanced unit test generation in C++.

\subsection{Results for Fine-Tuning (RQ-2)}

This section discusses the fine-tuning results for five different LLMs families. We hypothesize a PEFT model tuned on a task-specific demonstration data performs better than the corresponding base model for the task. Along the same line, we hypothesize a full-parameter fine-tuned model produces superior results than the corresponding PEFT counterpart. 

\subsubsection{Mistral-7B-Instruct-v0.2}

\begin{figure}[h!]
    \centering    \fbox{\includegraphics[width=0.8\textwidth]{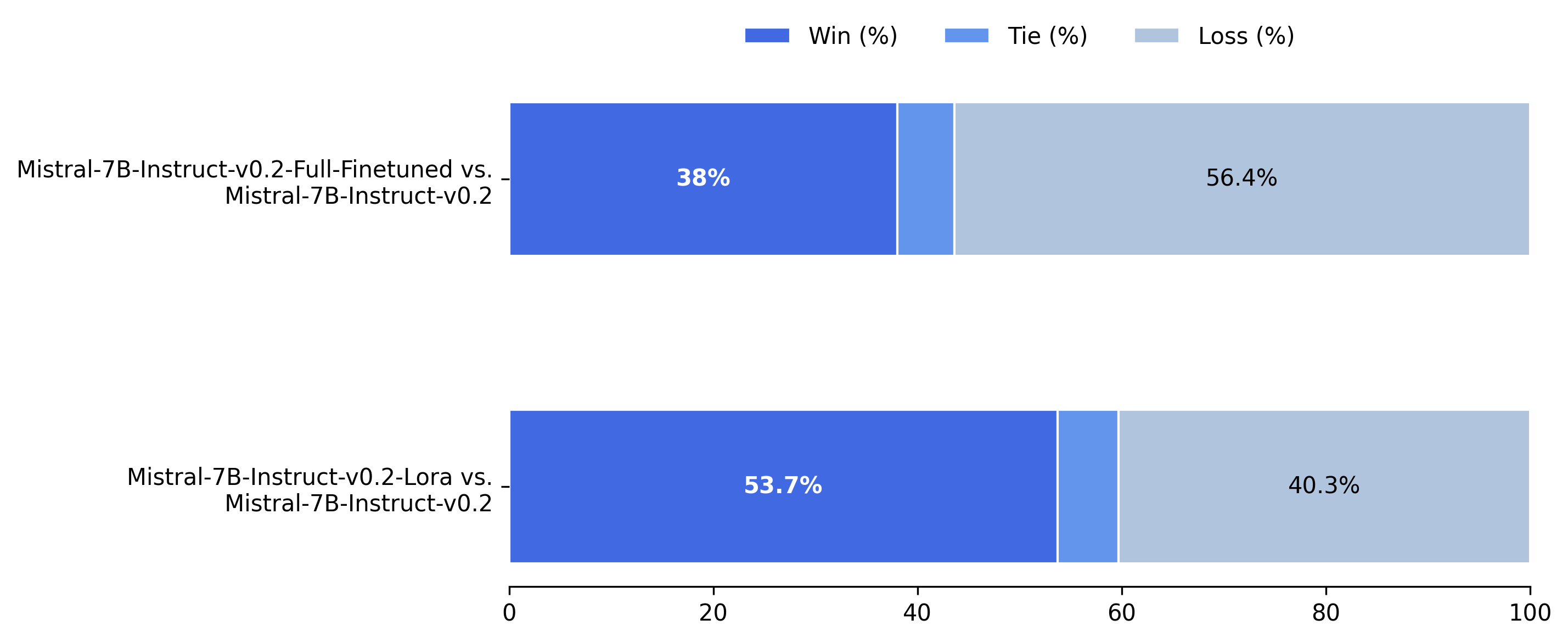}}
    \caption{Fine-tuning results for  Mistral-7B-Instruct-v0.2 \citep{mistral-v0.2}. The results corroborate with other general coding benchmarks. } 
    \label{fig:mistral-fine-tuning}
\end{figure}

Figure \ref{fig:mistral-fine-tuning} shows the win-rates for Lora-PEFT  Mistral-7B-Instruct-v0.2 vs Mistral-7B-Instruct-v0.2 and full-parameter fine-tuned Mistral-7B-Instruct-v0.2 vs Mistral-7B-Instruct-v0.2. It shows PEFT is working better than the base. But, quite surprisingly, the full-parameter model is performing poorly compared to the base model. This confounding observation can be explained by the MoE architecture \citep{zoph2022st}.

\subsubsection{TinyLlama}

\begin{figure}[h!]
    \centering    \fbox{\includegraphics[width=0.8\textwidth]{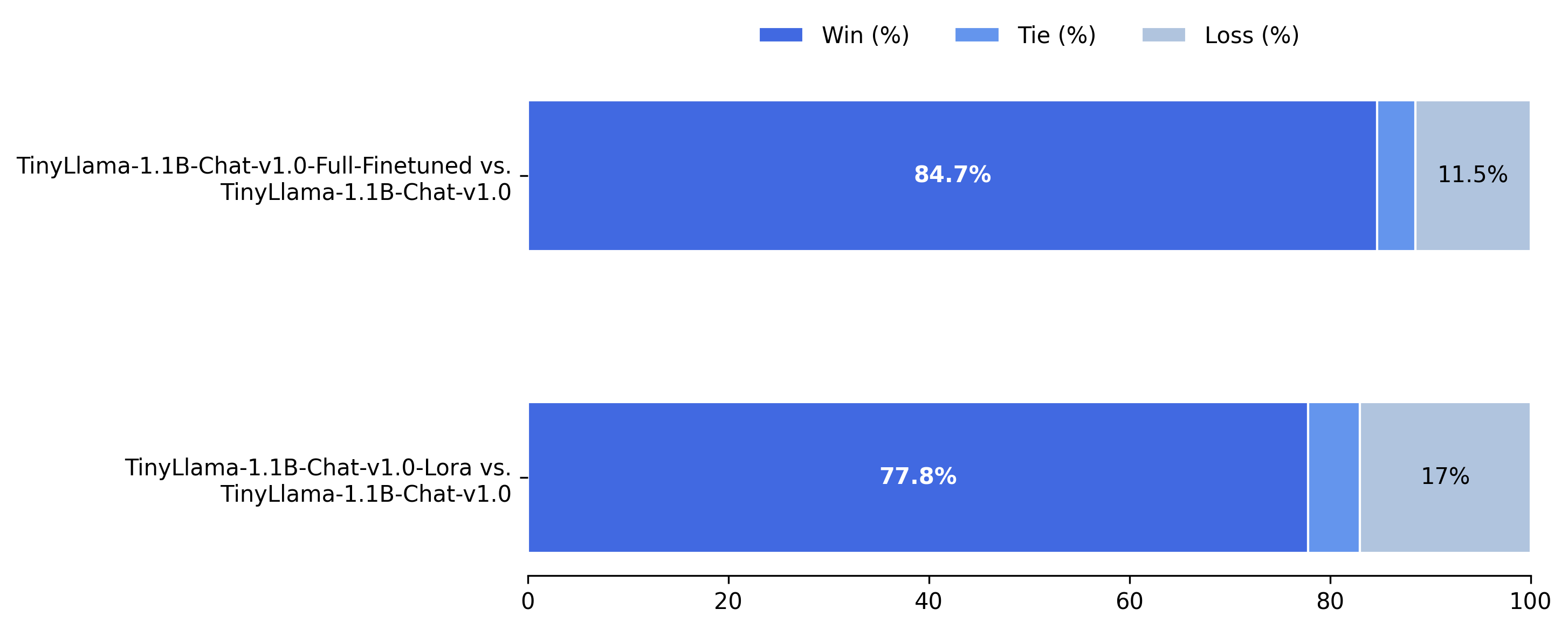}}
    \caption{Fine-tuning results for  TinyLlama \citep{tinyllama-1.1b}. The results corroborate with other general coding benchmarks. } 
    \label{fig:tiny-llama-fine-tuning}
\end{figure}

Figure \ref{fig:tiny-llama-fine-tuning} shows the win-rates for Lora-PEFT  TinyLlama-1.1B-Chat-v1.0 vs TinyLlama-1.1B-Chat-v1.0  and full-parameter fine-tuned TinyLlama-1.1B-Chat-v1.0 vs TinyLlama-1.1B-Chat-v1.0. It shows PEFT is working better than the base, winning 77.8\% times. With full-parameter fine-tuning the model performance improves further to 84.7\% w.r.t. the base. 

\subsubsection{CodeLlama}

\begin{figure}[h!]
    \centering    \fbox{\includegraphics[width=0.8\textwidth]{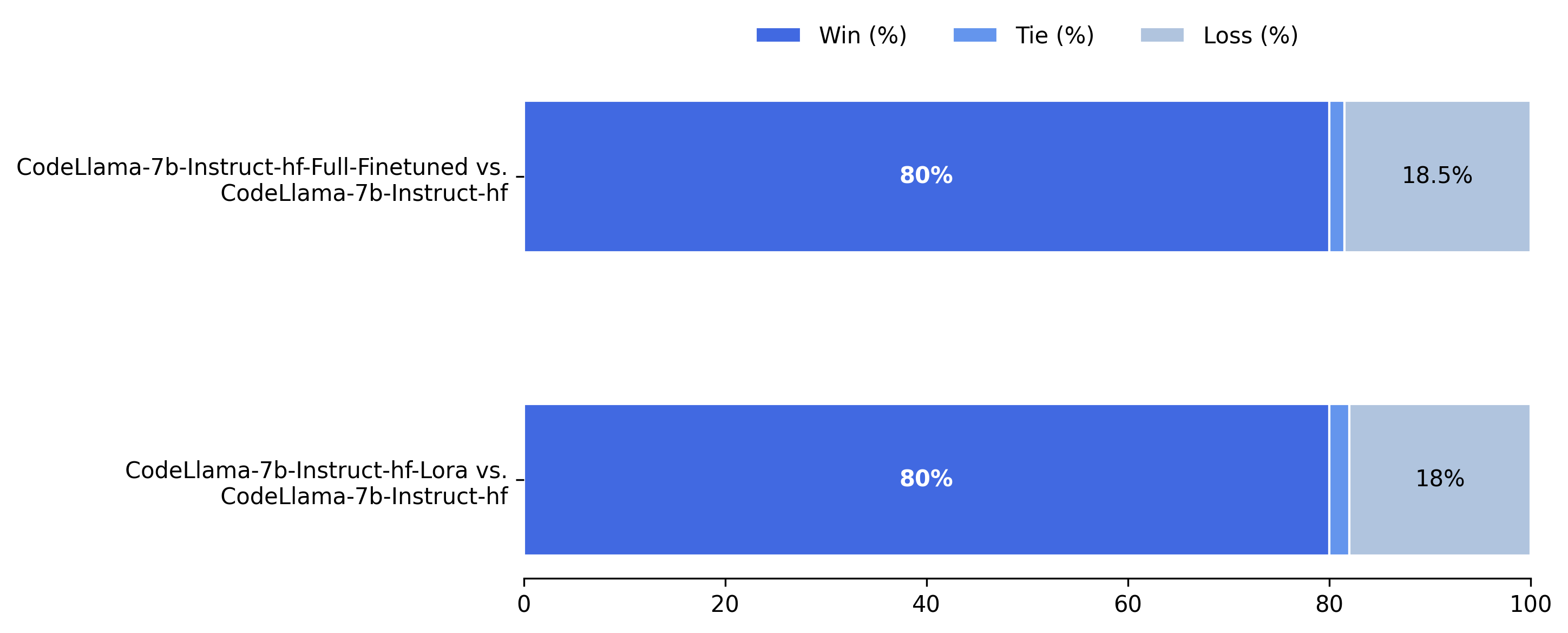}}
    \caption{Fine-tuning results for  CodeLlama-7B \citep{codellama-7b}   . The results corroborate with other general coding benchmarks. } 
    \label{fig:codellama-fine-tuning}
\end{figure}

Figure \ref{fig:codellama-fine-tuning} shows the win-rates for Lora-PEFT  CodeLlama-7B-Instruct-hf vs TinyLlama-1.1B-Chat-v1.0  and full-parameter fine-tuned CodeLlama-7B-Instruct-hf vs CodeLlama-7B-Instruct-hf. It shows both PEFT and full-parameter finetuning are performing on par with each other. This can be attributed to the relative strength of CodeLlama as a coding model, our hyper-parameter choice, $rank=8$, and relatively small data-size.

\subsubsection{Llama-3-8B}

\begin{figure}[h!]
    \centering    \fbox{\includegraphics[width=0.8\textwidth]{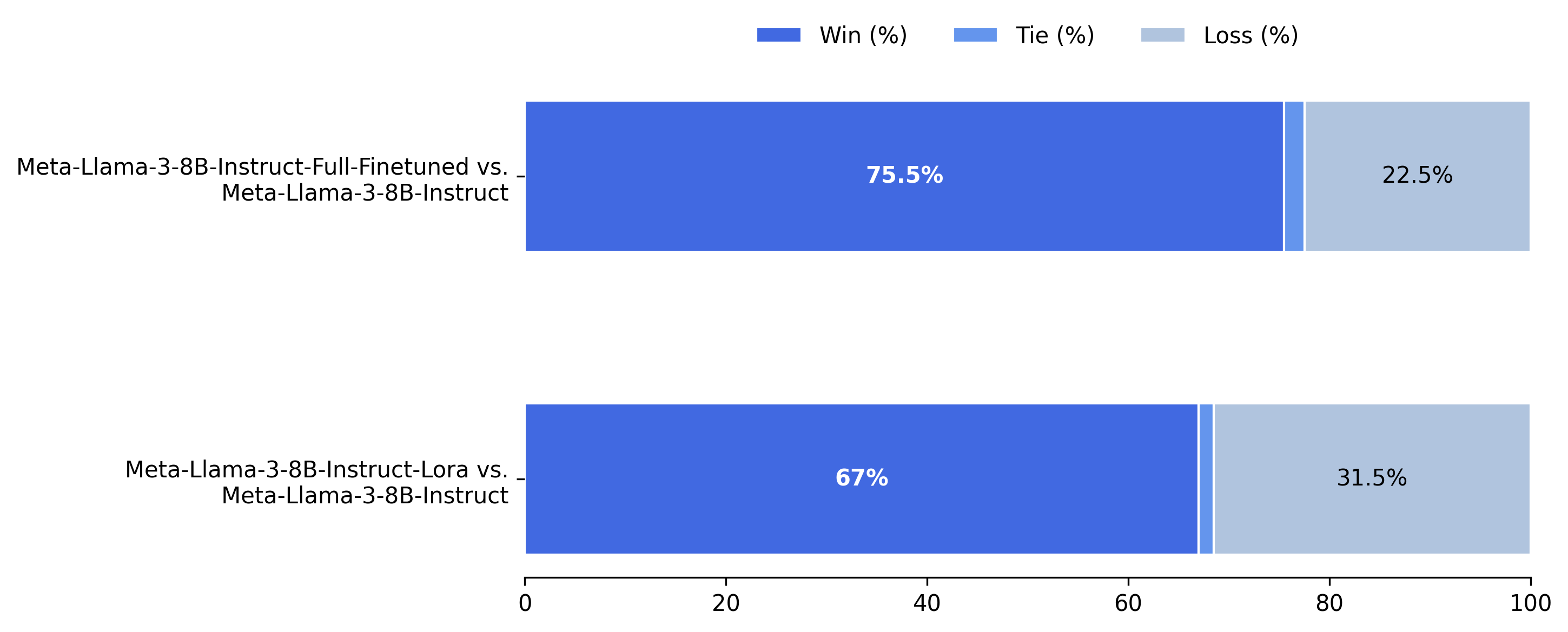}}
    \caption{Fine-tuning results for  Meta-Llama-3-8B-Instruct \citep{Llama-3-8B}   . The results corroborate with other general coding benchmarks. } 
    \label{fig:llama-3-fine-tuning}
\end{figure}

Figure \ref{fig:llama-3-fine-tuning} shows the win-rates for Lora-PEFT  Meta-Llama-3-8B-Instruct vs Meta-Llama-3-8B-Instruct  and full-parameter fine-tuned Meta-Llama-3-8B-Instruct vs Meta-Llama-3-8B-Instruct. It shows PEFT is working better than the base, winning 67\% times. With full-parameter fine-tuning the model performance improves further to 75.5\% w.r.t. the base. 

\subsubsection{Llama-3.1-8B}

\begin{figure}[h!]
    \centering    \fbox{\includegraphics[width=0.8\textwidth]{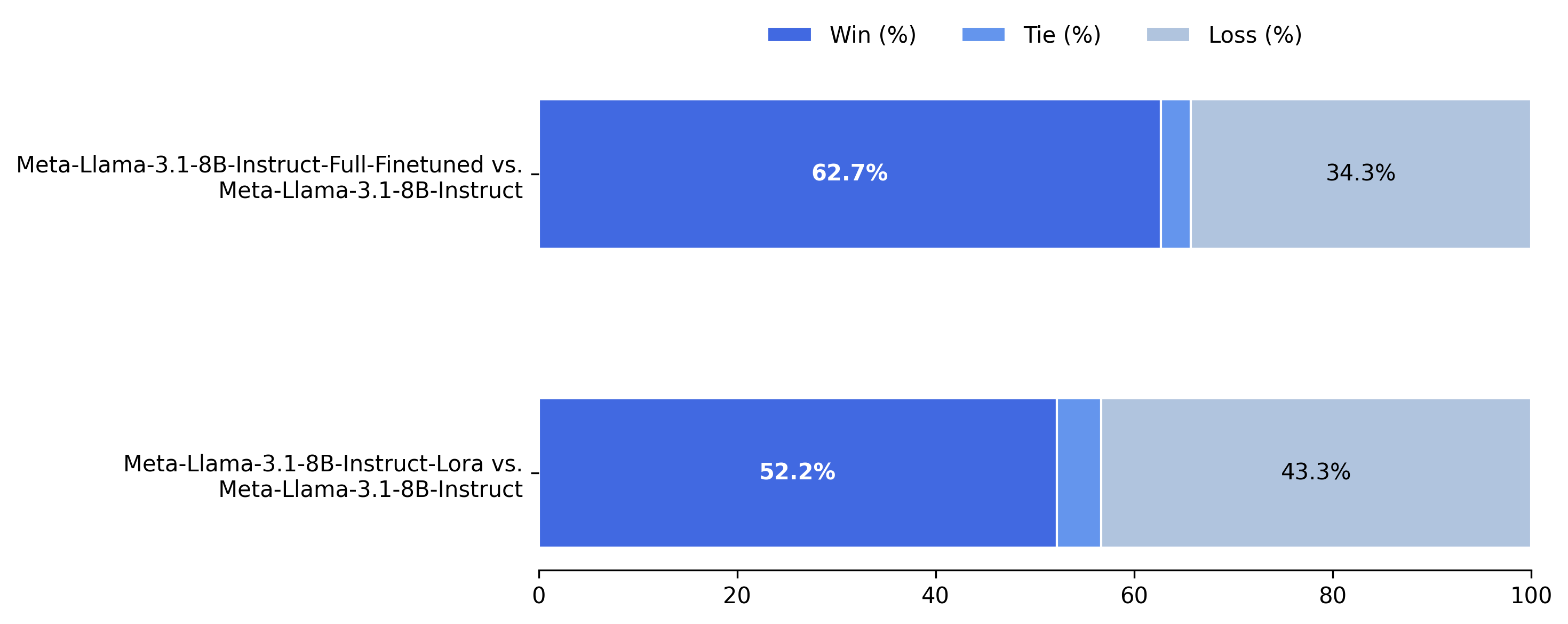}}
    \caption{Fine-tuning results for  Meta-Llama-3-8B-Instruct \citep{Llama-3.1-8B}   . The results corroborate with other general coding benchmarks. } 
    \label{fig:llama-3.1-fine-tuning}
\end{figure}

Figure \ref{fig:llama-3.1-fine-tuning} shows the win-rates for Lora-PEFT  Meta-Llama-3.1-8B-Instruct vs Meta-Llama-3.1-8B-Instruct  and full-parameter fine-tuned Meta-Llama-3.1-8B-Instruct vs Meta-Llama-3.1-8B-Instruct. It shows PEFT is working better than the base, winning 52.2\% times. With full-parameter fine-tuning the model performance improves further to 62.5\% w.r.t. the base. The relative improvement for Llama-3.1 is lower than Llama-3 can be explained by the superiority of former in the coding benchmarks \citep{dubey2024llama}.

\section{Conclusion}
In this work, we offer a C++ unit test benchmark, CPP-UT-Bench. We presented its scale and diversity across domains and features. We examined the effectiveness of CPP-UT-Bench for three different task scenarios: few-shot in-context learning, parameter-efficient fine-tuning (PEFT), and full-parameter fine-tuning for different LLM families. The patterns we discovered from our examinations corroborate with existing benchmarking standards. The resulting fine-tuned LLMs with CPP-UT-Bench show significant accuracy improvement compared to the base model. Therefore, we can claim the usability of CPP-UT-Bench as a benchmark dataset in C++ unit test generation with in-context learning and fine-tuning. For reproducibility, we will release our code. The future work will extend the scope to include alignment as well. 


\bibliography{neurips_2024}

\begin{thebibliography}{39}
\providecommand{\natexlab}[1]{#1}
\providecommand{\url}[1]{\texttt{#1}}
\expandafter\ifx\csname urlstyle\endcsname\relax
  \providecommand{\doi}[1]{doi: #1}\else
  \providecommand{\doi}{doi: \begingroup \urlstyle{rm}\Url}\fi

\bibitem[Abs()]{Abseil}
https://github.com/abseil/abseil-cpp.

\bibitem[CST()]{CST}
https://docs.sweep.dev/blogs/chunking-improvements.

\bibitem[Lla({\natexlab{a}})]{Llama-3-8B}
https://huggingface.co/meta-llama/Meta-Llama-3-8B, {\natexlab{a}}.

\bibitem[Lla({\natexlab{b}})]{Llama-3.1-8B}
https://huggingface.co/meta-llama/Meta-Llama-3.1-8B, {\natexlab{b}}.

\bibitem[cel()]{cel-cpp}
https://github.com/google/cel-cpp.

\bibitem[cod()]{codellama-7b}
https://huggingface.co/codellama/CodeLlama-7b-hf.

\bibitem[glo()]{glog}
https://github.com/google/glog?tab=readme-ov-file.

\bibitem[goo()]{google-test}
https://github.com/google/googletest.

\bibitem[lan()]{langsvr}
https://github.com/google/langsvr.

\bibitem[lev()]{leveldb}
https://github.com/google/leveldb.

\bibitem[lib({\natexlab{a}})]{libaddressinput}
https://github.com/google/libaddressinput, {\natexlab{a}}.

\bibitem[lib({\natexlab{b}})]{libphonenumber}
https://github.com/google/libphonenumber, {\natexlab{b}}.

\bibitem[lla({\natexlab{a}})]{llama-70b-awq}
https://huggingface.co/casperhansen/llama-3-70b-instruct-awq, {\natexlab{a}}.

\bibitem[lla({\natexlab{b}})]{llama-8b-awq}
https://huggingface.co/casperhansen/llama-3-8b-instruct-awq, {\natexlab{b}}.

\bibitem[mis({\natexlab{a}})]{mistral-v0.1}
https://huggingface.co/mistralai/Mistral-7B-v0.1, {\natexlab{a}}.

\bibitem[mis({\natexlab{b}})]{mistral-v0.2}
https://huggingface.co/mistralai/Mistral-7B-v0.2, {\natexlab{b}}.

\bibitem[phi({\natexlab{a}})]{phi-3-medium}
https://huggingface.co/microsoft/Phi-3-medium-128k-instruct, {\natexlab{a}}.

\bibitem[phi({\natexlab{b}})]{phi-3-small}
https://huggingface.co/microsoft/Phi-3-small-8k-instruct, {\natexlab{b}}.

\bibitem[pyt()]{pytorch}
https://github.com/pytorch/pytorch.

\bibitem[ten({\natexlab{a}})]{tensorflow}
https://github.com/tensorflow/tensorflow, {\natexlab{a}}.

\bibitem[ten({\natexlab{b}})]{tensorstore}
https://github.com/google/tensorstore, {\natexlab{b}}.

\bibitem[tin()]{tinyllama-1.1b}
https://huggingface.co/TinyLlama/TinyLlama-1.1B-Chat-v1.0.

\bibitem[tsl()]{tsl}
https://github.com/google/tsl.

\bibitem[Abdin et~al.(2024)Abdin, Jacobs, Awan, Aneja, Awadallah, Awadalla, Bach, Bahree, Bakhtiari, Behl, et~al.]{abdin2024phi}
Marah Abdin, Sam~Ade Jacobs, Ammar~Ahmad Awan, Jyoti Aneja, Ahmed Awadallah, Hany Awadalla, Nguyen Bach, Amit Bahree, Arash Bakhtiari, Harkirat Behl, et~al.
\newblock Phi-3 technical report: A highly capable language model locally on your phone.
\newblock \emph{arXiv preprint arXiv:2404.14219}, 2024.

\bibitem[Austin et~al.(2021)Austin, Odena, Nye, Bosma, Michalewski, Dohan, Jiang, Cai, Terry, Le, et~al.]{austin2021program}
Jacob Austin, Augustus Odena, Maxwell Nye, Maarten Bosma, Henryk Michalewski, David Dohan, Ellen Jiang, Carrie Cai, Michael Terry, Quoc Le, et~al.
\newblock Program synthesis with large language models.
\newblock \emph{arXiv preprint arXiv:2108.07732}, 2021.

\bibitem[Brown(2020)]{brown2020language}
Tom~B Brown.
\newblock Language models are few-shot learners.
\newblock \emph{arXiv preprint arXiv:2005.14165}, 2020.

\bibitem[Cassano et~al.(2023)Cassano, Gouwar, Nguyen, Nguyen, Phipps-Costin, Pinckney, Yee, Zi, Anderson, Feldman, et~al.]{cassano2023multipl}
Federico Cassano, John Gouwar, Daniel Nguyen, Sydney Nguyen, Luna Phipps-Costin, Donald Pinckney, Ming-Ho Yee, Yangtian Zi, Carolyn~Jane Anderson, Molly~Q Feldman, et~al.
\newblock Multipl-e: a scalable and polyglot approach to benchmarking neural code generation.
\newblock \emph{IEEE Transactions on Software Engineering}, 49\penalty0 (7):\penalty0 3675--3691, 2023.

\bibitem[Chen et~al.(2021)Chen, Tworek, Jun, Yuan, Pinto, Kaplan, Edwards, Burda, Joseph, Brockman, et~al.]{chen2021evaluating}
Mark Chen, Jerry Tworek, Heewoo Jun, Qiming Yuan, Henrique Ponde De~Oliveira Pinto, Jared Kaplan, Harri Edwards, Yuri Burda, Nicholas Joseph, Greg Brockman, et~al.
\newblock Evaluating large language models trained on code.
\newblock \emph{arXiv preprint arXiv:2107.03374}, 2021.

\bibitem[Dubey et~al.(2024)Dubey, Jauhri, Pandey, Kadian, Al-Dahle, Letman, Mathur, Schelten, Yang, Fan, et~al.]{dubey2024llama}
Abhimanyu Dubey, Abhinav Jauhri, Abhinav Pandey, Abhishek Kadian, Ahmad Al-Dahle, Aiesha Letman, Akhil Mathur, Alan Schelten, Amy Yang, Angela Fan, et~al.
\newblock The llama 3 herd of models.
\newblock \emph{arXiv preprint arXiv:2407.21783}, 2024.

\bibitem[Hu et~al.(2021)Hu, Shen, Wallis, Allen-Zhu, Li, Wang, Wang, and Chen]{hu2021lora}
Edward~J Hu, Yelong Shen, Phillip Wallis, Zeyuan Allen-Zhu, Yuanzhi Li, Shean Wang, Lu~Wang, and Weizhu Chen.
\newblock Lora: Low-rank adaptation of large language models.
\newblock \emph{arXiv preprint arXiv:2106.09685}, 2021.

\bibitem[Jiang et~al.(2023)Jiang, Sablayrolles, Mensch, Bamford, Chaplot, Casas, Bressand, Lengyel, Lample, Saulnier, et~al.]{jiang2023mistral}
Albert~Q Jiang, Alexandre Sablayrolles, Arthur Mensch, Chris Bamford, Devendra~Singh Chaplot, Diego de~las Casas, Florian Bressand, Gianna Lengyel, Guillaume Lample, Lucile Saulnier, et~al.
\newblock Mistral 7b.
\newblock \emph{arXiv preprint arXiv:2310.06825}, 2023.

\bibitem[Kiela et~al.(2021)Kiela, Bartolo, Nie, Kaushik, Geiger, Wu, Vidgen, Prasad, Singh, Ringshia, et~al.]{kiela2021dynabench}
Douwe Kiela, Max Bartolo, Yixin Nie, Divyansh Kaushik, Atticus Geiger, Zhengxuan Wu, Bertie Vidgen, Grusha Prasad, Amanpreet Singh, Pratik Ringshia, et~al.
\newblock Dynabench: Rethinking benchmarking in nlp.
\newblock \emph{arXiv preprint arXiv:2104.14337}, 2021.

\bibitem[Li(2008)]{li2008introduction}
M~Li.
\newblock An introduction to kolmogorov complexity and its applications, 2008.

\bibitem[Lopes and Hora(2022)]{lopes2022and}
Mateus Lopes and Andre Hora.
\newblock How and why we end up with complex methods: a multi-language study.
\newblock \emph{Empirical Software Engineering}, 27\penalty0 (5):\penalty0 115, 2022.

\bibitem[Lv et~al.(2024)Lv, Yang, Liu, Gao, Guo, and Qiu]{lv2024parameterfinetuninglargelanguage}
Kai Lv, Yuqing Yang, Tengxiao Liu, Qinghui Gao, Qipeng Guo, and Xipeng Qiu.
\newblock Full parameter fine-tuning for large language models with limited resources, 2024.
\newblock URL \url{https://arxiv.org/abs/2306.09782}.

\bibitem[Ott et~al.(2022)Ott, Barbosa-Silva, Blagec, Brauner, and Samwald]{ott2022mapping}
Simon Ott, Adriano Barbosa-Silva, Kathrin Blagec, Jan Brauner, and Matthias Samwald.
\newblock Mapping global dynamics of benchmark creation and saturation in artificial intelligence.
\newblock \emph{Nature Communications}, 13\penalty0 (1):\penalty0 6793, 2022.

\bibitem[Srivastava et~al.(2022)Srivastava, Rastogi, Rao, Shoeb, Abid, Fisch, Brown, Santoro, Gupta, Garriga-Alonso, et~al.]{srivastava2022beyond}
Aarohi Srivastava, Abhinav Rastogi, Abhishek Rao, Abu Awal~Md Shoeb, Abubakar Abid, Adam Fisch, Adam~R Brown, Adam Santoro, Aditya Gupta, Adri{\`a} Garriga-Alonso, et~al.
\newblock Beyond the imitation game: Quantifying and extrapolating the capabilities of language models.
\newblock \emph{arXiv preprint arXiv:2206.04615}, 2022.

\bibitem[Zheng et~al.(2023)Zheng, Chiang, Sheng, Zhuang, Wu, Zhuang, Lin, Li, Li, Xing, et~al.]{zheng2023judging}
Lianmin Zheng, Wei-Lin Chiang, Ying Sheng, Siyuan Zhuang, Zhanghao Wu, Yonghao Zhuang, Zi~Lin, Zhuohan Li, Dacheng Li, Eric Xing, et~al.
\newblock Judging llm-as-a-judge with mt-bench and chatbot arena.
\newblock \emph{Advances in Neural Information Processing Systems}, 36:\penalty0 46595--46623, 2023.

\bibitem[Zoph et~al.(2022)Zoph, Bello, Kumar, Du, Huang, Dean, Shazeer, and Fedus]{zoph2022st}
Barret Zoph, Irwan Bello, Sameer Kumar, Nan Du, Yanping Huang, Jeff Dean, Noam Shazeer, and William Fedus.
\newblock St-moe: Designing stable and transferable sparse expert models.
\newblock \emph{arXiv preprint arXiv:2202.08906}, 2022.

\end{thebibliography}
\bibliographystyle{plainnat}

\appendix

\section{Appendix / supplemental material}
\begin{table}[h!]
\label{tab:fine_tuning_hyperparameters}
\centering
\adjustbox{max width=\textwidth}{  
\begin{tabular}{|l|l|c|c|p{6cm}|}  
\hline
\textbf{Model Name}               & \textbf{PEFT Technique} & \textbf{Rank} & \textbf{Alpha} & \textbf{Layers Targeted}      \\ \hline
\textbf{Mistral-7B-Instruct-v0.2 }         & LoRA                    & 8             & 16             & up\_proj, o\_proj, gate\_proj, v\_proj, k\_proj, down\_proj, q\_proj \\ \hline
\textbf{Llama-3-8B-Instruct}             & LoRA                    & 8             & 16             & up\_proj, o\_proj, gate\_proj, v\_proj, k\_proj, down\_proj, q\_proj \\ \hline
\textbf{Llama-3.1-8B-Instruct }               & LoRA                    & 8             & 16             & q\_proj, v\_proj, k\_proj \\ \hline
\textbf{CodeLlama-7b-hf }        & LoRA                    & 8             & 16             & up\_proj, o\_proj, gate\_proj, v\_proj, k\_proj, down\_proj, q\_proj \\ \hline
\textbf{TinyLlama-1.1B-32k-Instruct}       & LoRA                    & 16            & 16             & up\_proj, o\_proj, gate\_proj, v\_proj, k\_proj, down\_proj, q\_proj \\ \hline

\end{tabular}
}
\vspace{0.1cm} 
\caption{Configurations for LoRA Fine-tuning of Different Models}
\end{table}

\section{Experimental Result Reproducibility}

To support the reproducibility of our experimental results, we provide links to the LoRA adapter weights and the fully finetuned model weights for each model used in our experiments. These resources allow other researchers to replicate the training procedures and fine-tuning outcomes presented in this paper.

The following table summarizes the models along with their corresponding weights:

\begin{table}[h!]
\centering
\begin{tabular}{|l|p{10cm}|} 
\hline
\textbf{Model Name} & \textbf{Links: LoRA Adapter Weights, Full Finetuned Model Weights} \\ \hline
\textbf{Mistral-7B-Instruct-v0.2} & \texttt{\small \url{https://huggingface.co/Nutanix/Mistral-7B-Instruct-v0.2\_cppunittest\_lora\_8\_alpha\_16}} \\ 
& \texttt{\small \url{https://huggingface.co/Nutanix/Mistral-7B-Instruct-v0.2\_cpp\_unit\_tests\_full\_finetuning\_class\_level}} \\ \hline
\textbf{Llama-3-8B-Instruct} & \texttt{\small \url{https://huggingface.co/Nutanix/Meta-Llama-3-8B-Instruct\_cppunittest\_lora\_8\_alpha\_16}} \\ 
& \texttt{\small \url{https://huggingface.co/Nutanix/Meta-Llama-3-8B-Instruct\_cppunittest\_full\_finetuning}} \\ \hline
\textbf{Llama-3.1-8B-Instruct} & \texttt{\small \url{https://huggingface.co/Nutanix/Meta-Llama-3.1-8B-Instruct\_cppunittest\_lora\_8\_alpha\_16}} \\ 
& \texttt{\small \url{https://huggingface.co/Nutanix/Meta-Llama-3.1-8B-Instruct\_cppunittest\_full\_finetuning}} \\ \hline
\textbf{CodeLlama-7b-hf} & \texttt{\small \url{https://huggingface.co/Nutanix/CodeLlama-7b-Instruct-hf\_cpp\_unit\_tests\_lora\_8\_alpha\_16\_class\_level}} \\ 
& \texttt{\small \url{https://huggingface.co/Nutanix/CodeLlama-7b-Instruct-hf\_cpp\_unit\_tests\_full\_finetuning\_class\_level}} \\ \hline
\textbf{TinyLlama-1.1B-32k-Instruct} & \texttt{\small \url{https://huggingface.co/Nutanix/TinyLlama-1.1B-32k-Instruct\_cppunittestprocessed\_lora\_16\_alpha\_16}} \\ 
& \texttt{\small \url{https://huggingface.co/Nutanix/TinyLlama-1.1B-32k-Instruct\_full\_finetuning}} \\ \hline
\end{tabular}
\vspace{0.5cm}
\caption{Links to Model Weights}
\end{table}

\newpage

\section{Unit Test Generation Prompt}

\begin{figure}[h!]
    \centering    \fbox{\includegraphics[width=0.8\textwidth]{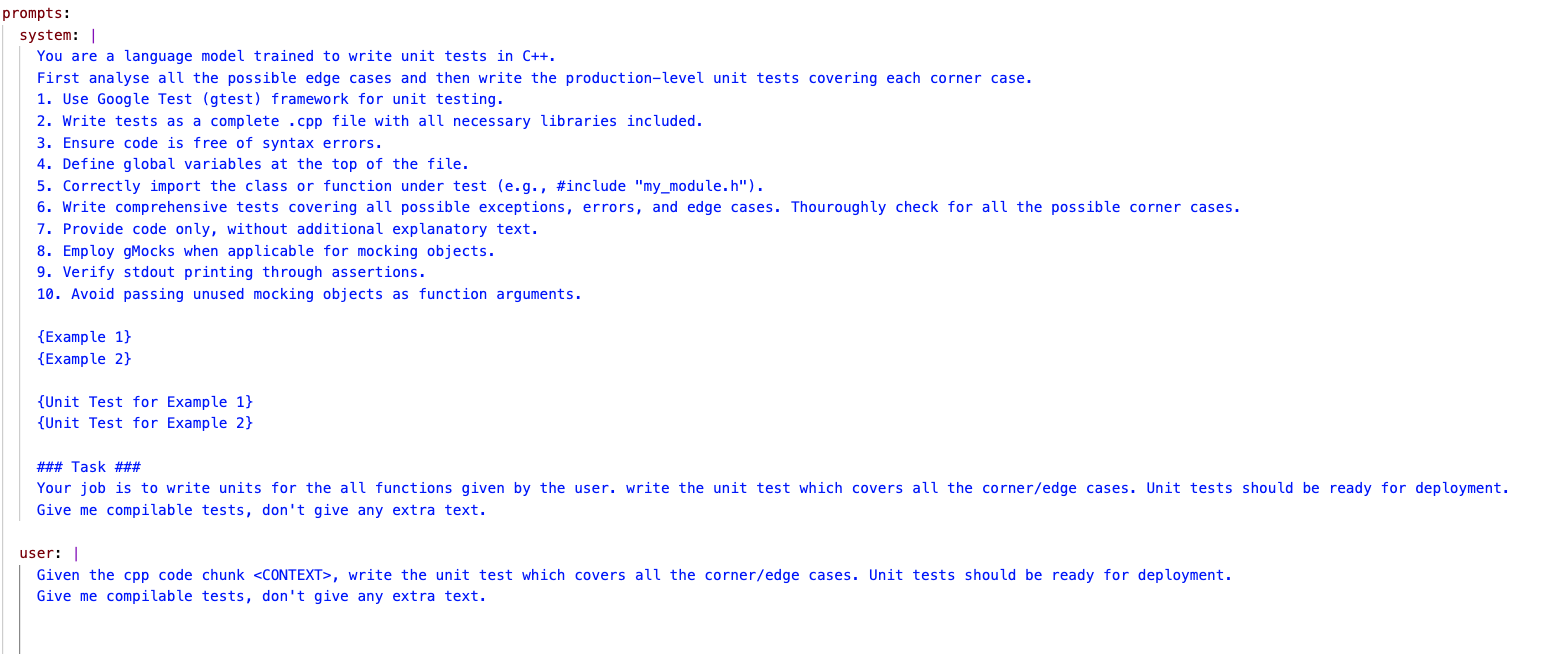}}
    \caption{Prompt template for the unit test generation in C++. } 
    \label{fig:prompt template}
\end{figure}

\end{document}